# Dynamic-nuclear-polarization-weighted spectroscopy of multi-spin electronic-nuclear clusters

Roberta Pigliapochi[1], Daniela Pagliero[1], Lisandro Buljubasich[3,4], Artur Lozovoi[1], Rodolfo H. Acosta[3,4], Pablo R. Zangara[3,4], and Carlos A. Meriles[1,2,*]

*[1]Department. of Physics, CUNY-City College of New York, New York, NY 10031, USA. [2]CUNY-Graduate Center, New York, NY 10016, USA. [3]Universidad Nacional de Córdoba, Facultad de Matemática, Astronomía, Física y Computación, Córdoba X5000HUA, Argentina. [4]CONICET, Instituto de Física Enrique Gaviola (IFEG), Córdoba X5000HUA, Argentina.*

Nuclear spins and paramagnetic centers in a solid randomly group to form clusters featuring nearly-degenerate, hybrid states whose dynamics are central to processes involving nuclear spin-lattice relaxation and diffusion. Their characterization, however, has proven notoriously difficult mostly due to their relative isolation and comparatively low concentration. Here, we combine field-cycling experiments, optical spin pumping, and variable radiofrequency (RF) excitation to probe transitions between hybrid multi-spin states formed by strongly coupled electronic and nuclear spins in diamond. Leveraging bulk nuclei as a collective time-integrating sensor, we probe the response of these spin clusters as we simultaneously vary the applied magnetic field and RF excitation to reconstruct multi-dimensional spectra. We uncover complex nuclear polarization patterns of alternating sign that we qualitatively capture through analytical and numerical modeling. Our results unambiguously expose the impact that strongly-hyperfine-coupled nuclei can have on the spin dynamics of the crystal, and inform future routes to spin cluster control and detection.

## I. INTRODUCTION

Nuclear spins proximal to paramagnetic impurities experience strong hyperfine couplings that often exceed their Zeeman interaction energies[1]. Because these couplings decay quickly with distance, nuclear spins can experience dramatically different magnetic resonance frequencies depending on their relative positions in the crystal lattice. This has led to the notion of a "spin-diffusion barrier", i.e., an imaginary region in space around the paramagnetic center where nuclei receive polarization efficiently but transfer it poorly due to the large energy mismatch with bulk spins[2-11]. This flip-flop quenching — and concomitant insulation of the core nuclear spins — is central to spin-lattice relaxation models as it impacts the rate of thermalization of bulk nuclear spins in a solid[12,13]. Analogously, it plays a key role in dynamic nuclear polarization[14,15] (DNP), presently attracting intense interest as a route to enhance the sensitivity of nuclear magnetic resonance[16,17].

Since there is no fundamental distinction between electronic and nuclear spins, similar ideas extend to the interaction of a paramagnetic impurity and an adjacent, strongly hyperfine-coupled nucleus, in the sense that spontaneous spin flip-flops are generically forbidden given the large energy differences between them (typically, a consequence of the large disparity between the electronic and nuclear gyromagnetic ratios). This situation changes, however, in the presence of multiple coupled paramagnetic centers provided the strength of this interaction is at least comparable to the nuclear spin energy. In this regime, a nuclear spin "flip" can be accompanied by a collective electronic "flop", i.e., a multi-spin reconfiguration whose final energy differs from the original one in an amount matching the nuclear energy splitting. Of particular interest is the case where one of the coupled paramagnetic centers has spin number greater than ½ because the above "matching" condition can be externally tuned via the proper selection of the applied magnetic field. The ensuing three-body electron–nuclear "cross-relaxation" process brings the above idea to its simplest conceptual realization[18-20].

Observing the many-body dynamics of these spin clusters via standard magnetic resonance techniques is difficult, not only because of the large frequency shifts separating core from bulk nuclei (typically in the tens or hundreds of MHz), but also owing to the relative amount of strongly hyperfine-coupled spins, invariably a small fraction of the total[4,7]. In principle, optically-detected magnetic resonance (ODMR) measurements — where a magneto-optically active color center is used as a local probe — can circumvent this problem[21-23], but because the sensitive volume is intrinsically limited to proximal nuclei, this technique can hardly gather information on the interplay between the internal spin dynamics of the cluster and spin diffusion into the bulk. Further, because pairs of interacting paramagnetic centers statistically form only in the limit of large concentrations, confocal microscopy methods — limited to probing multi-spin ensembles within the focal volume — are ill-suited to pick out contributions from individual spin clusters.

Here we combine optical electron spin pumping and field cycling to investigate the polarization dynamics of $^{13}C$ spins in diamond under continuous wave (cw) radiofrequency (RF) excitation. We focus on a crystal hosting a large concentration of nitrogen-vacancy (NV) centers and operate near "energy matching", i.e., within a magnetic field range where the separation between two levels in the NV ground state triplet nearly coincides with the electronic Zeeman splitting of neighboring spin-1/2 impurities[24-26]. By monitoring the bulk $^{13}$C NMR signal amplitude as we vary the operating magnetic field and RF frequency (much above the bulk $^{13}$C resonance), we unveil a rich set of hyperfine-coupling-sensitive $^{13}$C spectra, to the best of our knowledge, never seen before. We develop a model to compute the nuclear spin dynamics under continuous wave (cw) RF excitation, and show our observations collectively point to polarization processes mediated by a select group of strongly coupled spin clusters; this result highlights the system's ability to overcome the constraints imposed by the spin diffusion barrier and suggests intriguing routes to control and monitor strongly-hyperfine-coupled nuclei in an ensemble.

[*]Corresponding author. E-mail: cmeriles@ccny.cuny.edu.

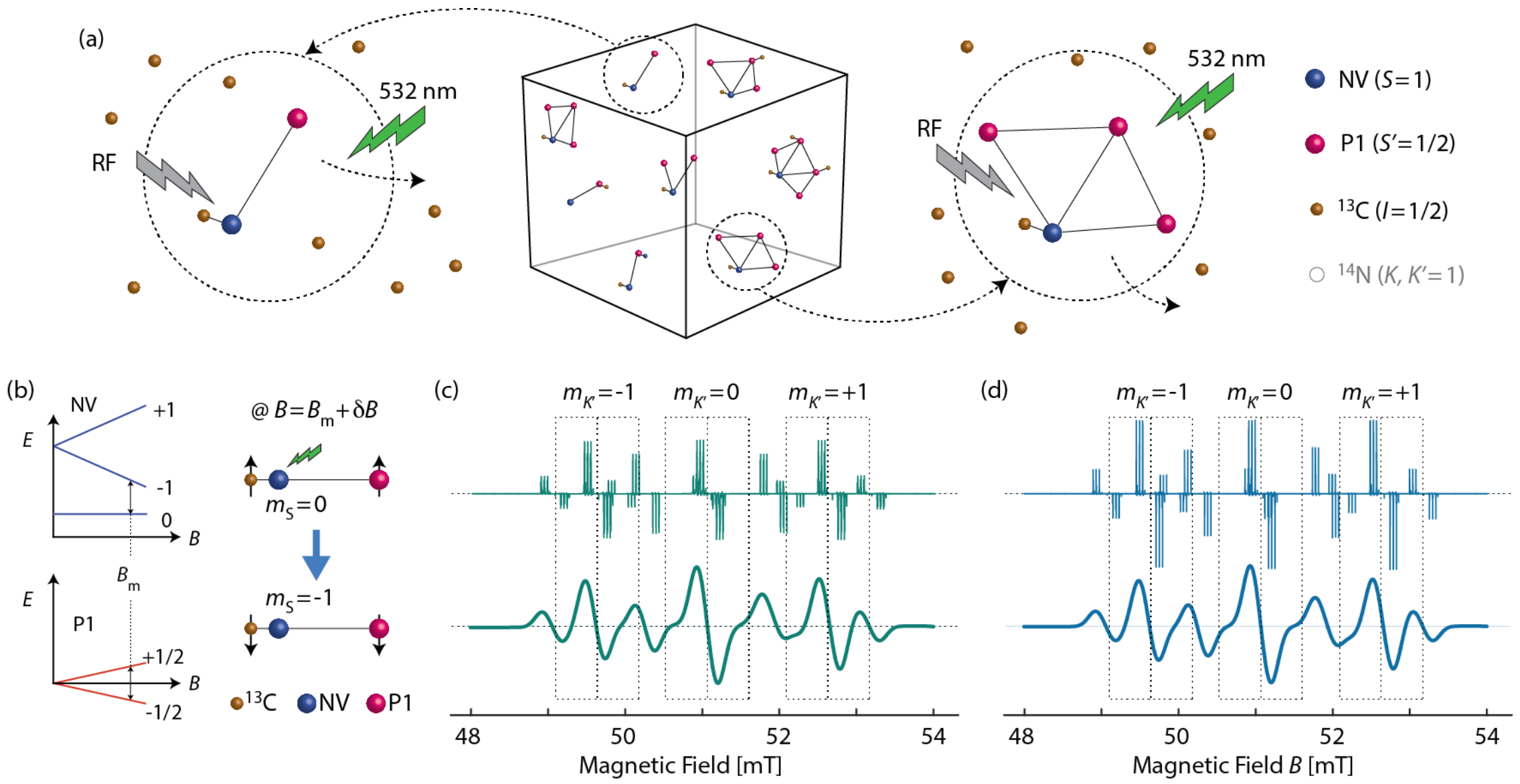

**Fig. 1: Optically pumped nuclear polarization as a multi-spin process.** (a) Randomly occurring paramagnetic centers (NVs, P1s) and $^{13}$C nuclei in the diamond lattice (as well as the $^{14}$N hosts of each point defect) can be grouped into hybrid clusters of strongly interacting spins dominating the generation of nuclear polarization during early stages of the DNP process. RF excitation resonant with hyperfine transitions in the cluster allows us to single out those nuclei most effectively contributing to the transfer of polarization to bulk carbons. (b) We work near $B_m$ where the P1 Zeeman splitting approximately matches one of the NV transitions (left energy diagrams). Slight detuning from $B_m$ provides the energy difference required to polarize the nuclear spin, as most clearly seen in a three-spin model (right schematics). (c) Calculated nuclear spin polarization pattern as a function of the applied magnetic field assuming the three-spin $^{13}$C–NV–P1 model in (b). Dashed boxes indicate sections of the pattern preferentially associated with each spin projection of the P1 $^{14}$N host. The lower trace derives from a convolution with a Gaussian, whose linewidth is normally chosen to attain best agreement with experiment. (d) Same as above but for the case of a spin cluster of the form $^{13}$C–NV–(P1)$_3$; the result can be hardly distinguished from that in (c).

## II. RESULTS

### II.1 Hyperfine spectroscopy via $^{13}$C DNP

Figure 1a lays out a visual representation of the spin system under investigation: Randomly distributed paramagnetic impurities in the form of NV centers and neutral substitutional nitrogen — also known as P1 centers — populate the diamond lattice with approximate concentrations of 10 and 70 ppm, respectively. For this rather high impurity content, fluctuations in the local density naturally lead to the formation of "spin clusters", i.e., groups of strongly interacting paramagnetic centers often comprising an NV and one (or more) P1 centers in close proximity.

Prior studies using the present crystal[26] and similar diamonds[20,25] have shown that green laser illumination — efficiently pumping the NV electronic spin— can induce bulk $^{13}$C spin polarization around 51.2 mT, the magnetic field where the energy separation between the $|m_S = 0\rangle$ and $|m_S = -1\rangle$ states of the NV nearly coincides with the $|m_{S'} = +1/2\rangle \leftrightarrow |m_{S'} = -1/2\rangle$ Zeeman splitting of the P1[24] (Fig. 1b). In particular, it has been shown one can semi-quantitatively reproduce the observed dependence of the bulk $^{13}$C polarization on the applied magnetic field by resorting to a four-spin model[25,26], which, besides the NV–P1 spin pair, includes a strongly coupled $^{13}$C and the $^{14}$N host of the P1 (Fig. 1c). Consistent with experiment (see below), the calculated polarization pattern roughly breaks down into three nearly equivalent sections, each associated with one of the possible nuclear spin projections of the $^{14}$N spin in the P1 (note that the hyperfine coupling between the NV electronic spin and its $^{14}$N host is much weaker and thus has a minor impact on the overall shape). This model, we warn, must be understood as a convenient (though crude) simplification, since disorder invariably leads to a statistical distribution of spin clusters whose average composition and configurational dispersion are largely a function of the crystal host preparation history. As a simple illustration, Fig. 1d shows that a $^{13}$C polarization pattern similar to that obtained in Fig. 1c can also be derived with identical assumptions starting from a different, more complex spin array. To account for this compositional heterogeneity, we employ throughout this article the more generic notion of a cluster even if, for simplicity, we limit our modeling to the $^{13}$C–NV–P1 set.

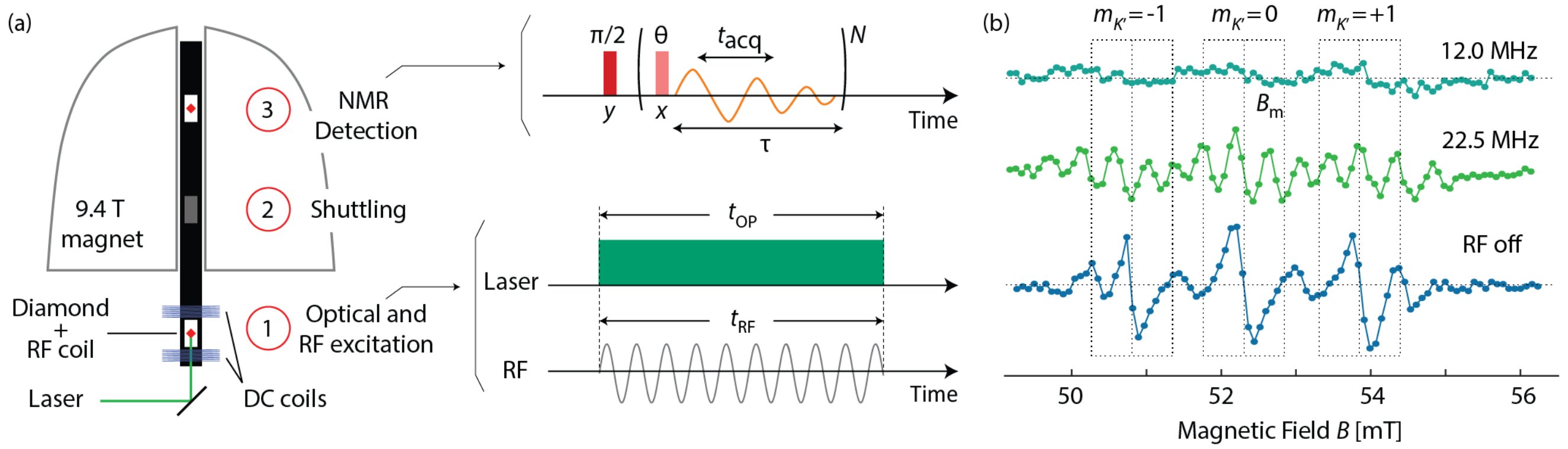

**Fig. 2: Dynamic nuclear polarization under continuous RF excitation.** (a) We implement a field cycling protocol where optical and RF excitation take place at low magnetic field $B$, which we vary in a vicinity of $B_m$; we detect the ensuing NMR signal after mechanical shuttling of the sample to a high-field magnet. (b) Experimental nuclear spin polarization patterns as a function of the applied magnetic field under RF excitation of variable frequency (see Appendix A for experimental details). Throughout these experiments the RF amplitude is $B_{RF} = 1$ mT; for comparison, the lower trace is the RF-free response. Dashed boxes have the same meaning as in Fig. 1.

To probe the dynamics of these spin clusters near $B_m$, we implement a protocol comprising simultaneous optical and RF excitation during an extended time window $t_{OP}$ (typically 10 s, see Fig. 2a as well as Appendix A). Intense green illumination spin-pumps the NVs into the $|m_S = 0\rangle$ state, while RF driving at variable frequencies allows us to probe transitions between hyperfine states associated to the multi-spin clusters governing the DNP process. As Fig. 2b illustrates, the impact of RF excitation on the observed signal — inductively detected upon sample shuttling to a high-field magnet, Fig. 2a — can be dramatic. For example, compared to the RF-free nuclear polarization pattern at variable magnetic field (lower trace in Fig. 2b, see also calculated trace in Fig. 1c), the $^{13}$C spin signal virtually vanishes under 12 MHz excitation, while 22.5 MHz RF leads to a strong, previously absent modulation (upper and middle traces in Fig. 2b, respectively). For reference, we emphasize that both frequencies are far removed from the bulk $^{13}$C Zeeman resonance (approximately 555 kHz at ~52 mT), which already hints at the important role of strongly hyperfine-coupled nuclei in the polarization transport process[7,26].

To better understand the dynamics at play, we focus on the central segment of the $^{13}$C spin polarization pattern (middle dashed box in Fig. 2b), and conduct systematic measurements of the induced nuclear magnetic resonance (NMR) signal as a function of the applied RF frequency and operating magnetic field. The range of fields closest to $B_{\rm m}$ belongs to the manifold formed by the $|m_{\rm K'} = 0\rangle$ projection of the $^{14}$N nuclear host of the P1: Limiting our study to this group of transitions is a reasonable simplification, as we anticipate equivalent dependences in the subsets associated with the $|m_{\rm K'} = \pm 1\rangle$ projections (left and right dashed boxes in Fig. 2b). Note that while joint spin-flip processes also involving $^{14}$N nuclei are known to occur[25,26], these processes are comparatively less probable and thus play a less prominent role in defining the $^{13}$C field polarization pattern.

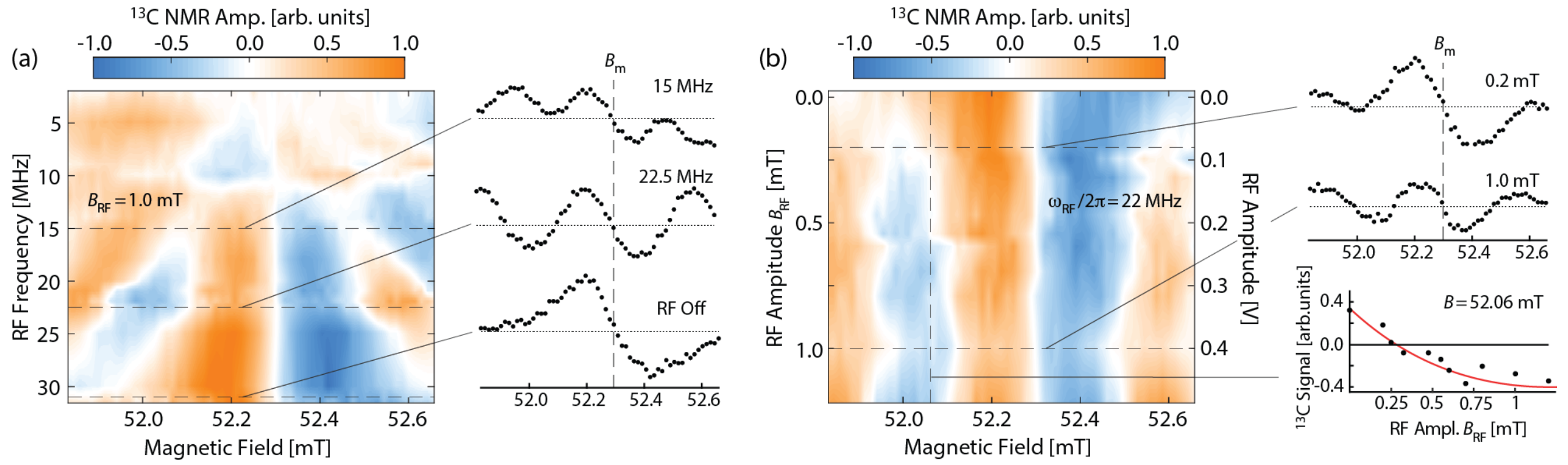

**Fig. 3: Hyperfine spectroscopy of strongly coupled nuclei.** (a) Observed $^{13}$C NMR signal as a function of the applied magnetic field and RF frequency. Insert plots on the right are horizontal cross sections of the two-dimensional plot at the indicated frequencies. For comparison, the lower plot shows the RF-free pattern (last row in the two-dimensional plot). We observe a correspondence between the applied RF frequency and the magnetic field range where the $^{13}$C signal reverses sign. Throughout these measurements the RF amplitude is 1 mT. (b) $^{13}$C NMR signal as a function of the RF amplitude and applied magnetic field under 22 MHz RF excitation. The upper inserts on the right-hand side are horizontal cross sections of the measured response at two different RF amplitudes; the lower insert plot is a vertical cross section at ~52.06 mT, where we observe maximal RF-induced change of the NMR signal amplitude; the solid line is a guide to the eye.

Figure 3a captures our observations throughout the $|m_{\rm K'} = 0\rangle$ range of magnetic fields as we vary the RF frequency $\omega_{\rm RF}$ over a broad bandwidth, from 3 to 30 MHz. Importantly, these frequencies are much greater than the $^{13}$C Larmor frequency, meaning that RF excitation directly addresses hyperfine-coupled nuclei presumably isolated from bulk spins by a large diffusion barrier. Remarkably, we observe a strong response that extends over the entire RF range we probe. The center of symmetry in the pattern — shifted from the theoretical value of 51.2 mT due to slight misalignment of the NV axis[26,27] — flags the "matching" field $B_{\rm m}$ in these experiments. As $\omega_{\rm RF}$ grows, regions of inverted $^{13}$C NMR signal emerge, gradually displacing to the outer sections of the inspected magnetic field range. That the observed signal can change its sign upon RF excitation — rather than, e.g., simply decrease to zero — is itself intriguing as it points to spin processes involving not just the carbon nuclei (see below). Working at the magnetic field where the effect is greatest, we find the degree of inversion depends smoothly on the applied RF amplitude (Fig. 3b), progressively saturating as we reach the conditions in Fig. 3a.

Overall, these results can be seen as a form of nuclear spin hyperfine spectroscopy, valuable in that they allow us to directly gauge the influence of alternative channels on the generation of nuclear polarization. This is possible because the signal we observe — encoded in the level of polarization of bulk nuclei — depends not only on the number of $^{13}$C spins resonant with a given RF frequency but, more importantly, on how effectively they enable the flow of spin order from its source — in this case, the NV — to the bulk of the crystal. Note that since bulk nuclear spins feature longer spin-lattice relaxation times, they act here as a memory, recording the selective effect of the RF on a much smaller group of nuclear spins as a time-integrated change in the bulk spin magnetization. Despite the huge frequency disparity between the spin resonances of bulk nuclei and those we address, the results in Fig. 3 make it clear that strongly coupled $^{13}$C spins must play a key role at early stages of the DNP process (in the sense that one would expect no effect unless nuclei proximal to paramagnetic centers communicate efficiently with bulk spins). Unfortunately, the complex spectral signatures we observe make the in-depth understanding required to deconvolve these polarization channels far from straightforward; we tackle this problem immediately below.

### II.2 Modeling DNP under continuous RF excitation

To interpret our observations, we consider a $^{13}$C–NV–P1 spin cluster (SC) Hamiltonian in a static (but variable) magnetic field $B$, namely,

$$H_{\rm SC} = H_{\rm NV} + H_{\rm P1} + H_{\rm C} + H_{HF} + H_{\rm d}. \qquad (1)$$

In Eq. (1), $H_{\rm NV} = DS_z^2 + |\gamma_{\rm e}|BS_z$ contains the NV crystal field and Zeeman interactions, $H_{\rm P1} = |\gamma_{\rm e}|BS'_z$ and $H_{\rm C} = -\gamma_{\rm n}BI_z$ respectively represent the P1 and $^{13}$C Zeeman couplings, $H_{\rm HF} = A_{zz}S_zI_z + A_{zx}S_zI_x$ is the $^{13}$C–NV hyperfine interaction, $H_{\rm d} = \mathcal{J}_{\rm d}S_zS'_z - (3\tilde{\mathcal{J}}_{\rm d}/4)(S_+S'_+ + S_-S'_-)$ is the NV–P1 dipolar contribution with coupling constants $\mathcal{J}_{\rm d}$ and $\tilde{\mathcal{J}}_{\rm d}$. In the presence of RF, the Hamiltonian

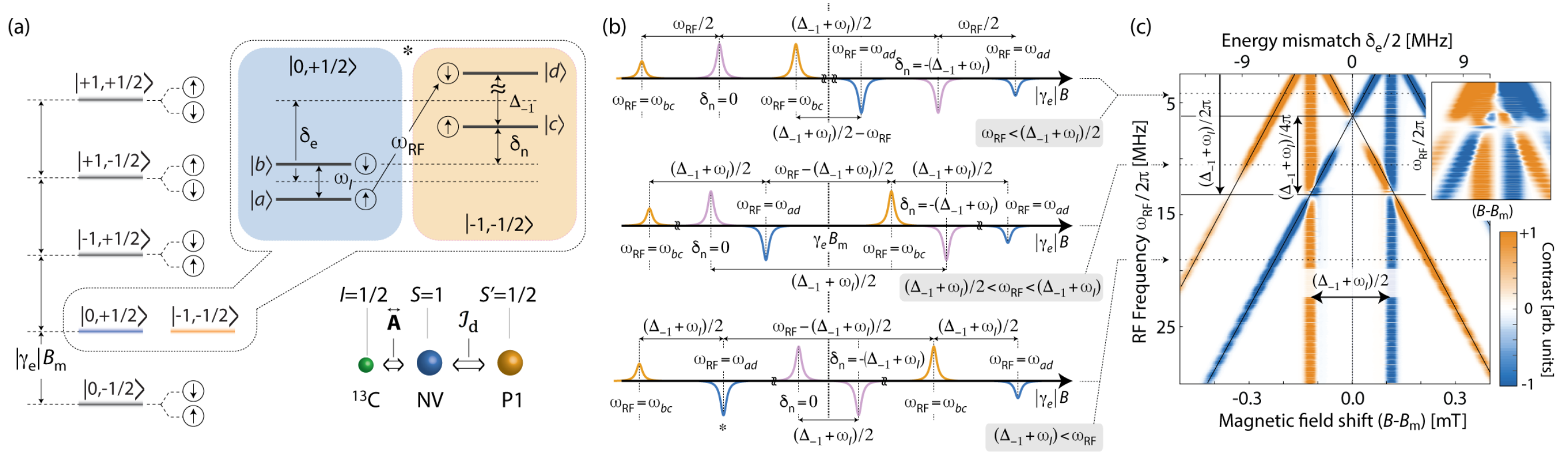

**Fig. 4: Modeling spin dynamics under cw RF excitation.** (a) Schematic of the energy diagram for a $^{13}$C–NV–P1 triad near the matching condition. The right-hand side insert is a zoomed view of the nearly degenerate state manifold within the dashed square. (b) Anticipated nuclear polarization pattern as a function of the electronic Zeeman frequency $|\gamma_{\rm e}|B$. The upper, middle, and lower traces respectively describe the cases where $\omega_{\rm RF} < (\Delta_{-1} + \omega_I)$, $(\Delta_{-1} + \omega_I)/2 < \omega_{\rm RF} < (\Delta_{-1} + \omega_I)$, and $(\Delta_{-1} + \omega_I) < \omega_{\rm RF}$. The asterisk in the left-hand peak of the lower schematics refers to the conditions of the insert in (a); faint purple traces indicate nuclear polarization stemming from level crossing in the lab frame (i.e., in the absence of RF). Note we use different horizontal scales in each case. (c) Calculated $^{13}$C polarization as a function of the applied magnetic field and excitation frequency. For these calculations, we assume $A_{zz} = A_{zx} = 11.7$ MHz and $\mathcal{J}_{\rm d} = \tilde{\mathcal{J}}_{\rm d} = 250$ kHz; the RF amplitude is $B_{\rm RF} = 2$ mT. (Upper right-hand insert) To facilitate comparison with experiment, we convolve the resulting pattern at each frequency with a broad Gaussian.

must be supplemented with the term

$$H_{\mathrm{RF}} = (|\gamma_{\mathrm{e}}|S_x + |\gamma_{\mathrm{e}}|S'_x - \gamma_{\mathrm{n}} I_x) B_{\mathrm{RF}} \cos(\omega_{\mathrm{RF}} t), \quad (2)$$

which describes the time-dependent coupling with an RF field of amplitude $B_{\mathrm{RF}}$ and frequency $\omega_{\mathrm{RF}}$. In the above formulas, $D = 2\pi \times 2.87$ GHz is the NV crystal field, $\gamma_{\mathrm{e}} = -2\pi \times 28.025$ GHz T$^{-1}$ and $\gamma_{\mathrm{n}} = 2\pi \times 10.71$ MHz T$^{-1}$ respectively denote the electron and $^{13}$C gyromagnetic ratios, we assume $\hbar = 1$, and use the standard notation for spin operators.

Figure 4a lays out the energy diagram of the $^{13}$C–NV–P1 cluster: In this representation, energy matching at $B_{\mathrm{m}}$ amounts to a degeneracy between the $|m_S = 0, m_{S'} = +1/2\rangle$ and $|m_S = -1, m_{S'} = -1/2\rangle$ states. Since green light spin-pumps the NV into $|m_S = 0\rangle$, nuclear polarization stems from a cross-relaxation process that invariably starts in the $|m_S = 0, m_{S'} = +1/2\rangle$ manifold; hyperfine couplings — exclusively active for $|m_S = -1\rangle$ — shift the condition for energy matching away from $B_{\mathrm{m}}$, hence leading to nuclear spin polarization of one sign or the other as one varies the magnetic field[26]. By the same token, RF excitation can effectively create a (rotating frame) degeneracy between levels in either manifold, thus opening alternative polarization transfer channels. Therefore, we must interpret the observed NMR signal at the chosen frequency and magnetic field as an incoherent superposition of contributions from complementary spin clusters, namely, those away from RF resonance but energy matched in the lab frame, and those whose contribution at the working magnetic field is made possible only through the presence of RF.

Deriving a formal rotating-frame description that accurately captures the impact of RF on the system dynamics is difficult because degeneracies lead to electronic/nuclear state hybridization with the consequence that virtually no transition between states in the manifold can be a priori excluded, especially given the varying nature of the applied magnetic field. Further, it is precisely due to this hybridization that one cannot generically ignore the electronic term in $H_{\mathrm{RF}}$ despite the gigantic mismatch between $\omega_{\mathrm{RF}}$ (in the range 3–30 MHz) and the electronic Zeeman transition frequencies (in the GHz range for a ~52 mT magnetic field).

Despite the above caveats, however, it is possible to anticipate the sign and qualitative shape of the nuclear polarization pattern at variable field[28], which we lay out in Fig. 4b for distinct RF frequency ranges in a scale governed by the hyperfine-induced splitting $\Delta_{-1} = (A_{zx}^2 + (A_{zz} + \omega_I)^2)^{1/2}$, where we defined the $^{13}$C Larmor frequency $\omega_I = \gamma_{\mathrm{n}} B$ (see insert in Fig. 4a). Assuming for simplicity optical initialization into the $|0, +1/2\rangle$ manifold, the sign and relative position of each peak in the pattern follow from considering the nuclear spin character of the states connected by the RF field (expressed as an "up" or "down" arrow). For example, for the energy alignment sketched in the insert to Fig. 4a (corresponding to the case where $\Delta_{-1} < \omega_{\mathrm{RF}}$), a simple exercise shows that all polarization peaks must alternate sign (lower trace in Fig. 4b). Note that the polarization transfer efficiency — qualitatively reflected by the amplitude of the corresponding peak in the spectrum — is expected to decrease as $B$ departs from $B_{\mathrm{m}}$ because hybridization between the $|0, +1/2\rangle$ and $|-1, -1/2\rangle$ states gradually vanishes away from matching.

More rigorously, we first leverage the Hamiltonian in Eq. (1) to numerically calculate the $^{13}$C polarization pattern for a representative cluster. To take into account the impact of RF on the system dynamics, we transform $H_{\mathrm{SC}}$ to a frame rotating at the excitation frequency, and enforce selection rules that preserve pairs of temporally averaged states whose energy difference is comparable to $\omega_{\mathrm{RF}}$ for a given applied magnetic field and coupling parameter set (see Appendix B). Figure 4c shows the results assuming a $^{13}$C–NV–P1 cluster with dipolar (hyperfine) coupling constants $\mathcal{J}_{\mathrm{d}} = \tilde{\mathcal{J}}_{\mathrm{d}} = 250$ kHz ($A_{zz} = A_{zx} = 11.7$ MHz, roughly corresponding to the extrema in the RF-free polarization pattern, Fig. 3a). In the low frequency range, we find that satellites either share the sign of the RF-free peak (if $\omega_{RF} \lesssim \Delta_{-1}/2$, upper trace in Fig. 4b), or show opposite polarization (when $\Delta_{-1}/2 \lesssim \omega_{RF} \lesssim \Delta_{-1}$, middle trace in Fig. 4b), consistent with the schematic. Above ~13 MHz — i.e., in the limit $\omega_{RF} \gtrsim \Delta_{-1}$ — we find a series of positive and negative peaks, reminiscent of that observed experimentally at medium to high frequencies (see traces at 22.5 MHz in Figs. 3a and 2b).

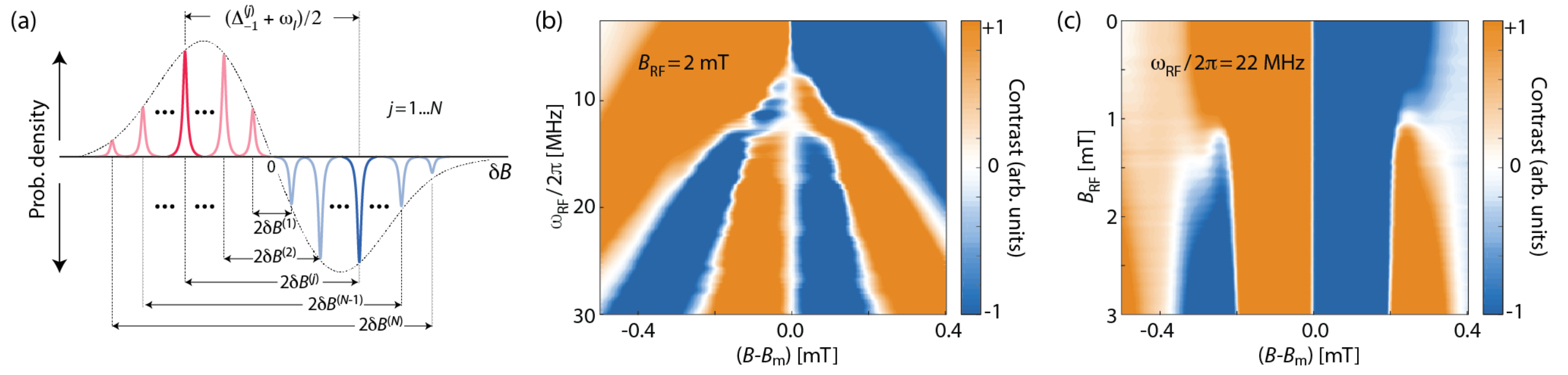

**Fig. 5: The impact of coupling heterogeneity.** (a) Either half of the RF-free nuclear polarization pattern can be interpreted as a probability distribution reflecting on the number and polarization efficiency of spin clusters $j = \{1 \dots N\}$, each featuring a hyperfine coupling $\left|\Delta_{-1}^{(j)}\right| \approx \left|4\gamma_{\mathrm{e}}\, \delta B^{(j)}\right|$ evenly distributed across the relevant field shift range $|\delta B| \equiv |B - B_{\mathrm{m}}|$. (b) Calculated nuclear polarization as a function of the RF frequency and applied magnetic field assuming an RF amplitude $B_{\mathrm{RF}} = 2$ mT. For these calculations, we use a collection of $N = 22$ three-spin clusters distributed across a detuning range $[0 \cdots 0.4]$ mT, and assign weights chosen to match the experimental RF-free polarization pattern as shown in (a). (c) Same as in (b) but as a function of $B_{\mathrm{RF}}$ and field detuning $(B - B_{\mathrm{m}})$ for $\omega_{\mathrm{RF}}/2\pi = 22$ MHz.

The findings above suggest the system polarization dynamics as a whole can be crudely described as governed by spin clusters featuring hyperfine couplings of order ~10 MHz. In particular, we regain some of the key experimental features if we convolve the calculated signal with a Gaussian whose broadening (0.05 mT) was chosen to best match the data (upper right-hand insert in Fig. 4c). With only one hyperfine coupling being considered, however, this approach is necessarily rudimentary, hence raising the question as to whether the moderate agreement we find can be improved by more carefully weighing in contributions from clusters with different couplings. Furthermore, reproducing the observations in Fig. 3 requires we simultaneously consider contributions due to spin clusters whose transitions are detuned from a given RF frequency but energy-matched at the applied magnetic field.

To model the impact of disorder in the crystal lattice, we first note that the signal contribution stemming from a "matched" cluster (i.e., the cluster satisfying the condition $|\Delta_{-1}| \approx |4\gamma_e(B - B_m)|$ at a given field $B$, see Fig. 5a) derives not only from the calculated $^{13}$C polarization but also from the transport efficiency and the cluster's relative abundance, all of which combines to yield the measured RF-free pattern. Correspondingly, we reinterpret this latter pattern — or, more precisely, each half of it — as a probability distribution, which we subsequently leverage to weigh in contributions from individual spin clusters $j = \{1 \dots N\}$ in a discrete collection with varying hyperfine couplings $\Delta_{-1}^{(j)}$. Figures 5b and 5c show the results as a function of $B$, as well as the RF frequency and amplitude. Comparison with the experimental observations in Figs. 3a and 3b shows that the agreement — though reasonable — remains moderate. This is particularly the case for Fig. 5b where we calculate diagonal bands of inverted polarization whose relative positions and amplitudes differ from those observed. Further, our calculations predict strong RF-induced modulations even at 30 MHz, not seen experimentally.

Some of these problems can be mitigated by lifting the constraint on the relative cluster weights. In particular, we find closer agreement with our observations when we emphasize contributions from clusters featuring stronger hyperfine couplings; we show two illustrations in Figs. 6a and 6b for alternative cluster histograms. Despite a slight improvement, however, the end result is not entirely satisfactory, especially when we note that a change in the cluster weight assignments necessarily has a direct impact on the calculated RF-free pattern.

In retrospect, the lack of quantitative agreement should not be surprising given the disparity between the three-spin-model we use herein and the broad cluster heterogeneity intrinsic to any realistic crystal (Fig. 1a). Along these lines, the notion of a hyperfine-coupling continuum reaching up to ~47 MHz (necessary to explain the observed width in the RF-free pattern, see lower trace in the insert to Fig. 3a) is inconsistent with the discrete nature of the diamond lattice (featuring large gaps in the set of allowed values above ~15 MHz[29-31]). On the other hand, differences in the RF amplitude required to induce inversion of the $^{13}$C signal — noticeably stronger in our calculations as compared to the observations in Fig. 3b — likely arises from the effective enhancement of the gyromagnetic ratio of strongly-coupled $^{13}$C nuclei, as observed in optically-detected magnetic resonance spectra[32] (for simplicity not considered in our model Hamiltonian).

## III. DISCUSSION

In summary, the combined use of color center spin pumping, magnetic field tuning, and RF excitation far from the nuclear Larmor frequency allows us to probe a broad set of strongly coupled spin clusters integrating two (or more) paramagnetic centers and one (or more) hyperfine-coupled nuclei. Systematic studies as a function of the applied field and RF excitation reveal a rich, though complex, response, which we qualitatively capture through a model that simultaneously includes the impact of hyperfine heterogeneity and RF excitation near the electronic spin level anti-crossing at $B_m$.

Admittedly, the notion of a crystalline host as a collection of isolated clusters — here exploited to facilitate numerical computations in otherwise intractable ensembles — must be understood as a starting working framework because weaker interactions between clusters must necessarily be present to transport polarization from strongly-hyperfine-coupled to bulk nuclei[7,33]. By construction, therefore, our approach is insensitive to clusters that do not communicate efficiently with the spin environment and hence little (if anything) can be said about their number and composition. On the other hand, the observation of efficient bulk nuclear spin polarization stemming from *some* strongly coupled clusters highlights the role of electron-electron interactions in accelerating nuclear spin diffusion between nuclei with dissimilar resonance frequencies. An immediate corollary is that the notion of a "spin diffusion barrier" must be applied carefully as it clearly breaks down in systems such as the present one[4-7]. Something similar can be said about nuclear s pin-lattice relaxation models describing depolarization as the result of interactions with isolated electronic spin fluctuators.

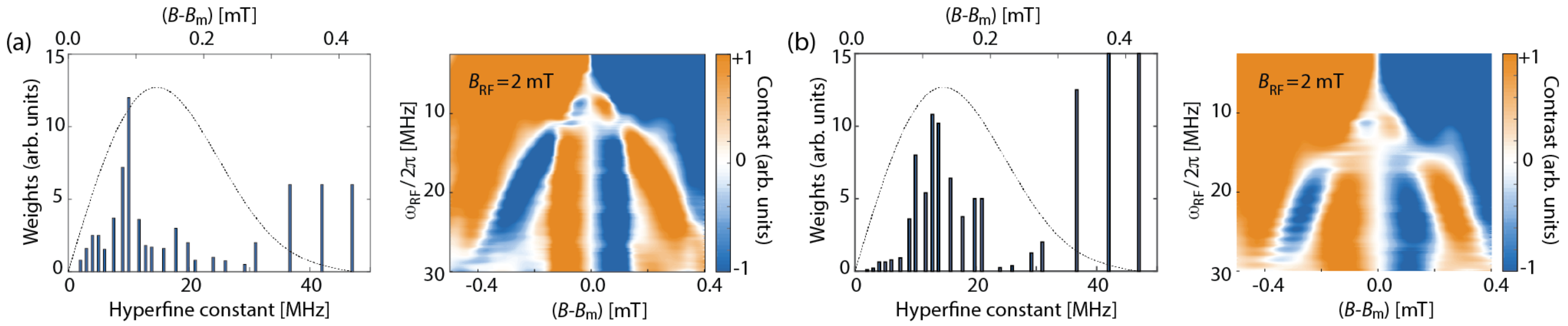

**Fig. 6: The impact of coupling heterogeneity.** (a) Weight histogram (left) and calculated polarization pattern (right) assuming large contributions from strongly hyperfine-coupled clusters. The dotted line on the left-hand plot is the hyperfine-coupling distribution function adapted to match the RF-free pattern. (b) Same as in (a) but for an alternative weight histogram.

The ability to read the action of RF excitation through the polarization of bulk nuclei makes it arguably possible to interrogate spin clusters in a crystal — and, more generally, the many-body network they form — in ways thus far unexplored. For example, unlike the experiments above — where the bulk signal expresses the steady state polarization of cluster $^{13}$C spins under continuous optical and RF excitation — one could imagine separating cluster initialization and control through time-resolved schemes featuring multiple repeating units, each integrating laser and RF pulses as well as periods of free evolution. Of particular interest is the regime where the separation $t_{\mathrm{W}}$ between consecutive repetitions is longer than the time required to erase the cluster memory (in turn, defined by the inverse spin diffusion rate $(\Gamma_{\mathrm{d}})^{-1}$). In this limit, every unit in the temporal train becomes nearly independent, and the observed bulk polarization can hence be seen as the integrated result of multiple identical experiments of duration $t_{\mathrm{C}}$, each of which can be adapted to probe the cluster in arbitrary ways. Ensuring the condition $\Gamma_{\mathrm{d}} t_{\mathrm{C}} \gtrsim 1$, however, is difficult in practice because the finite nuclear spin-lattice relaxation time $T_{1\mathrm{n}}$ imposes an upper bound on the number $K$ of pulse train units contributing to the integrated signal. Therefore, pulsed protocols featuring long wait times become impractical if $K$ is much lower than the minimum number of repeats required to imprint an observable bulk magnetization (the case herein). This limitation, nonetheless, could be circumvented at lower temperatures, where nuclear spin-lattice relaxation times become sufficiently long.

Gaining temporal control over these cross-polarization processes is important, not only from a fundamental standpoint but also as a practical route to new DNP strategies. For example, we have theoretically shown that controlled thermal jumps combined with RF excitation of electron/nuclear spin transitions close to (but detuned from) energy matching should lead to efficient dynamic nuclear polarization without the need for microwave (MW)[34]. This approach is attractive in that it potentially circumvents some of the physical and practical hurdles complicating the implementation of DNP techniques at high magnetic fields (typically, 1-2 T or greater, where a gyrotron is required for efficient MW generation). Additional work, however, will be needed to gain a fuller understanding of the RF impact on these hybrid transitions, particularly in cases as the present one, where the composition and coupling strength in the clusters is broadly heterogeneous.

## ACKNOWLEDGMENTS

The work of R.P. and C.A.M. was supported by the U.S. Department of Energy (DOE), Office of Science, Basic Energy Sciences (BES) under Award BES-DE-SC0020638. D.P. and A.L. acknowledges support from the National Science Foundation through grant NSF-2203904. R.P., D.P. and C.A.M. also acknowledge access to the facilities and research infrastructure of the NSF CREST IDEALS, grant number NSF-2112550. L.B., R.H.A., and P.R.Z. acknowledge financial support from CONICET (PIP-111122013010074 6CO), SeCyT-UNC (33620180100154CB).

## APENDIX A: Sample and experimental setup

The experimental conditions resemble those presented in detail in previous work[26]. Briefly, we use in our present experiments an HPHT [100] diamond with a nitrogen content of 70 ppm. The preparation protocol comprised high-energy electron irradiation (7 MeV at a dose of $10^{18}$ cm$^{-2}$) and subsequent thermal annealing (2 h at 700°C) resulting in an NV$^-$ concentration of about 10 ppm. The crystal — with dimensions 3.2 × 3.2 × 0.3 mm$^3$ — is attached to a sapphire holder, itself part of a custom-made NMR probe head. The diamond support is attached to a cogwheel system, to allow for alignment of the [111] crystal axis and the magnetic field in such a way that the external magnetic field nearly coincides with one of the NV axes. The diamond used in this work has been previously characterized[26], and detailed knowledge of the orientation of the NV axes was available. Inside the probe head, the diamond is in contact with a 3-mm diameter, 3-loop coil ('RF coil' hereafter) used for RF excitation, and, on the other side, a 3-mm diameter 6-loop coil ('NMR coil' hereafter) used for $^{13}$C NMR excitation/detection, which we carry out in a 9.4 T NMR spectrometer. The 'NMR coil' is connected to a pair of variable capacitors to allow for tuning and matching, which are also integrated within the NMR probe-head. The 'RF coil' is connected to an external amplifier, to control the frequency and power of the RF excitation. Unlike in prior work[28], this system allows us to attain high-power RF excitation over a broad frequency bandwidth.

The NMR probe head initially sits vertically outside and below the NMR magnet bore, experiencing a stray field of approximately 52 mT. There, a pair of electromagnetic coils connected to a power supply (Instek PSM-6003 operated in current control mode) allows us to vary the applied magnetic field over a $\pm 5$ mT range. Via the delivered current, the effective magnetic field experienced by the sample is swept and tuned to the hyperpolarization condition of need, as and tuned to the hyperpolarization condition of need, as described in the main text. While exposed to this (low) magnetic field, the diamond is illuminated with a 532-nm laser beam. The on/off pulsing of the laser is controlled using an acousto-optic modulator (AOM), which limits the beam power to approximately 500 mW at the sample. The beam size at the diamond surface is optimized through an optical lens placed at about 15 cm from the sample to obtain a beam size radius of approximately 1 mm.

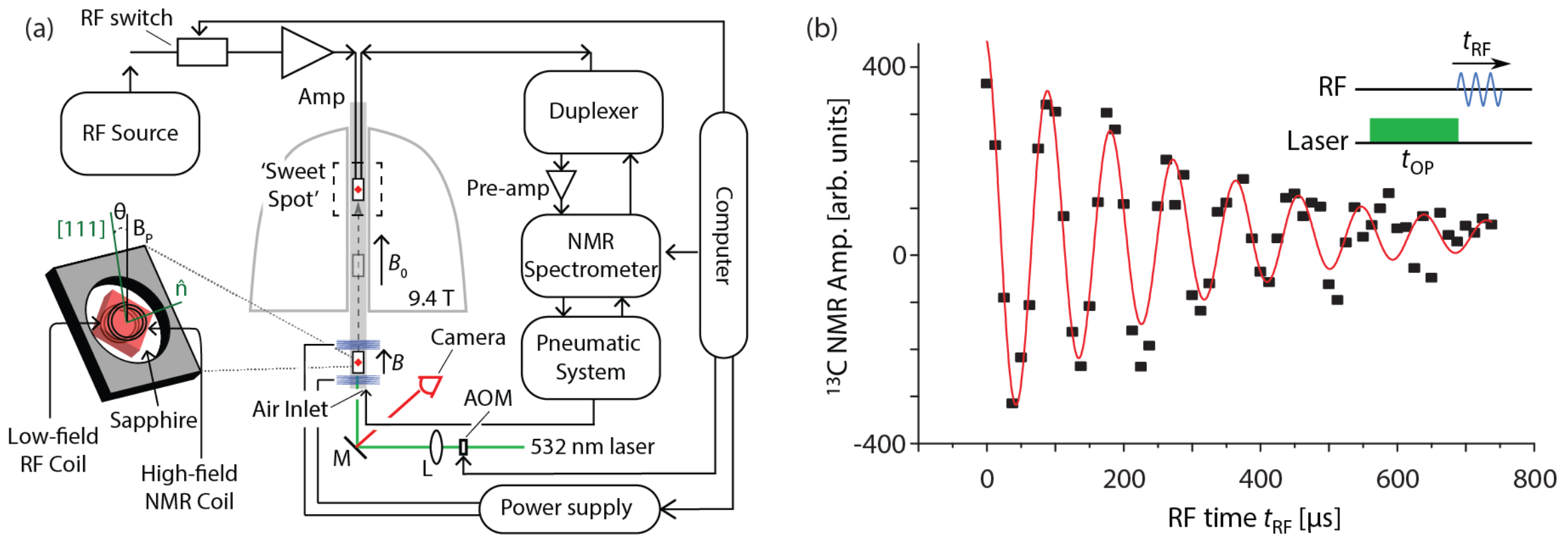

**Fig. 7: Experimental details.** (a) Schematics of the experimental setup. The diamond (red square) sits on a sapphire crystal holder, itself part of a custom-made NMR probe head. The diamond support is aligned so that the [111] crystal axis is nearly parallel to the external field $B$. In the insert drawing, $\hat{\mathbf{n}}$ denotes the diamond surface normal; θ represents a misalignment of the diamond leading to a deviation from the 51.2 mT as the matching polarization field. The diamond sits between a 3-mm diameter 3-loop coil ('RF coil') used for RF excitation at low field, and a 3-mm diameter 6-loop coil ('NMR coil') used for $^{13}$C NMR excitation/detection in a 9.4 T NMR magnet. The 'NMR coil' is connected to a pair of variable capacitors (not shown) to allow for tuning and matching, which are also integrated within the NMR probe-head. The 'RF coil' is connected to an external source, to control the RF frequency and power. In its resting position under the stray field, a pair of electromagnetic coils (shown in blue) allow us to vary the magnetic field amplitude; 532 nm laser excitation takes place while under this field. The on/off laser pulsing is controlled using an acousto-optic modulator (AOM), which determines a laser beam power of approximately 500 mW at the sample. We use an optical lens (L) ~15 cm from the sample to optimize the excitation spot. After optical illumination and RF excitation at low field, the probe head is shuttled up inside the NMR magnet using a custom-made pneumatic system, and held at a stable 'sweet' spot for NMR excitation/detection via the 'NMR coil'. The timings of shuttling and NMR excitation/detection are controlled using TTL commands from a master computer; we use a camera to monitor the sample and the laser beam spot on its surface at regular intervals. (b) Measurement of the Rabi frequency. The sequence, schematized on the top right, consists of a 10-second laser illumination of the diamond, followed by an RF pulse (-6 dBm) of variable duration resonant with the $^{13}$C Larmor frequency (558 kHz); NMR detection is carried out after mechanical shuttling to 9.6 T (not shown). Throughout this measurement, $B$~52 mT for maximum $^{13}$C NMR signal and the number of repeats per point is 16; the solid red line is a fit of a damped sinusoidal.

Depending on the chosen spin preparation protocol, a radio-frequency wave is generated through the 'RF coil' of varying frequency, power, and time duration. Following optical and RF spin manipulation at low field, the probe head is sent up inside a 9 T NMR magnet bore using a custom-made pneumatic shuttling system, and held at a stable 'sweet' spot — i.e., where the magnetic field is maximum and optimally homogeneous — for inductive NMR detection. The timings of shuttling and NMR spectrometer triggering are controlled using TTL commands. In order to ensure stability in the diamond orientation over consecutive runs, the sample and the laser beam spot on its surface are monitored using an external camera (see system schematics in Fig. 7a).

For high-field $^{13}$C NMR excitation/detection, we implement an adaptation of the multi-pulse sequence in Ref. [35]. The protocol consists of an initial 90° pulse (22 μs), followed by a train of $M$ equally spaced, 90-degree-phase-shifted pulses of fixed flip angle θ (31.4 μs), equivalent to a spin rotation by an amount approximately equal to 128°. The free-induction-decay we collect emerges from stroboscopic acquisition between consecutive θ pulses. We set the period in the θ-pulse train to 91 μs (i.e., a 60 μs inter-pulse interval) for a total number of 4000 pulses. Each collection is repeated and averaged over 16 times.

## APENDIX B: Spin modeling

Deriving an effective, time-independent Hamiltonian $H_{\text{eff}}$ that captures the action of the RF field is difficult because electron/nuclear hybridization and degeneracy near a level anti-crossing activate otherwise-forbidden, multi-spin transitions between varying pairs of levels. Deciding which coupling terms can and cannot be truncated in the resulting time-dependent matrix representing the Hamiltonian is not at all apparent hence making a careful preparatory analysis mandatory, even if, as in the present case, the end goal is to attain a numerical result (the system complexity is sufficient to make computing times easily diverge). Hybridization and degeneracy depend, of course, on the applied magnetic field — here varying across the full range of level crossings — with the consequence that obtaining generic expressions of $H_{\text{eff}}$ becomes especially challenging. We implement, therefore, an approach integrating theory and numerical modeling featuring the following sequence of steps[28]:

(*i*) For a given magnetic field $B$ and RF frequency $\omega_{\mathrm{RF}}$ — both of which take arbitrary but fixed values — we transform the spin cluster Hamiltonian $H_{\mathrm{SC}}$ in Eq. (1) to diagonal form $T^{-1}H_{\mathrm{SC}}T$, where $T$ denotes the transformation matrix into its eigenbasis.

(*ii*) Arranging the eigenstates by their energy in increasing order, the upper third corresponds to states with $m_S = +1$; these do not play any role in the system dynamics and can be ignored. The rest belongs to the $m_S = -1$ and/or $m_S = 0$ manifolds, where no associations or classifications are a priori possible owing to the dependence on $B$ and the proximity to the matching field $B_{\mathrm{m}}$.

(*iii*) We define an operator with non-zero entries ($= \omega_{\mathrm{RF}}$) only in the diagonal, medium third of the eigenbasis,

$$\mathbb{W} = \begin{bmatrix} 0 & & & & & & & \\ & \ldots & & & & & & \\ & & 0 & & & & & \\ & & & \omega_{\mathrm{RF}} & & & & \\ & & & & \ldots & & & \\ & & & & & \omega_{\mathrm{RF}} & & \\ & & & & & & 0 & \\ & & & & & & & \ldots \\ & & & & & & & & 0 \end{bmatrix} \quad \text{(B1)}$$

Notice that $\mathbb{W}$ is the same for any magnetic field, irrespective of how $B$ compares to $B_{\mathrm{m}}$. The corresponding transformation into the effective rotating frame is given by the operator $R = \exp\{-i\mathbb{W}t\}$.

(*iv*) We transform the diagonalized spin-cluster Hamiltonian into the rotating frame and define $\widetilde{H}_{\mathrm{SC}} = RTH_{\mathrm{SC}}T^{-1}R^{-1}$. Notice that $\widetilde{H}_{\mathrm{SC}}$ remains diagonal and time independent.

(*v*) We compute the off-diagonal matrix elements coming from $H_{\mathrm{RF}}$. To this end, we first transform into the eigenbasis defined in step (*i*), and then into the effective rotating frame, i.e., we calculate $\widetilde{H}_{\mathrm{RF}} = RTH_{\mathrm{RF}}T^{-1}R^{-1}$. Subsequently, we set to zero all diagonal elements in $\widetilde{H}_{\mathrm{RF}}$, and average the time dependence in the off-diagonal elements by resorting to the identity

$$\langle \exp\{\pm i\omega_{\mathrm{RF}}t\}\cos(\omega_{\mathrm{RF}}t)\rangle_{\frac{2\pi}{\omega_{\mathrm{RF}}}} = \frac{1}{2}. \quad \text{(B2)}$$

(*vi*) The effective, time-independent Hamiltonian is therefore given by

$$H_{\mathrm{eff}} = \widetilde{H}_{\mathrm{SC}} - \mathbb{W} + \widetilde{H}_{\mathrm{RF}}\,. \quad \text{(B3)}$$

We emphasize that the first two contributions in the above formula are diagonal and the term $-\mathbb{W}$ corresponds to the standard $-i\dot{R}R^{-1}$. The term $\widetilde{H}_{\mathrm{RF}}$ is purely non-diagonal and yields the transition matrix elements between eigenstates induced by the RF driving.

(*vii*) Starting from $H_{\mathrm{eff}}$, we numerically calculate the $^{13}$C polarization in a given $^{13}$C-NV-P1 cluster of fixed hyperfine coupling constant for varying magnetic field $B$ and RF frequency $\omega_{RF}$. For example, the results in Fig. 4c reproduce the response for a total of 30,000 magnetic field intensities and 300 RF frequencies, assuming a common RF amplitude of 2 mT (slightly above the experimental value and where we numerically attain nearly optimal signal inversion, see Fig. 5c). The same calculation was performed independently for $N = 22$ different hyperfine coupling constants ranging from 2 MHz to 47 MHz. To capture the coupling heterogeneity in the system, we represent the total $^{13}$C polarization as a linear combination of individual spin clusters with hyperfine coupling $\Delta_{-1}^{(j)}$ where $j = \{1 \ldots N\}$, and assign weights $\eta_{\Delta}$. Figure 5b displays the result for a weight set that matches the experimental RF-free profile, as discussed in the main text. Another group of simulations with variable $\Delta_{-1}^{(j)}$ were implemented with RF frequency fixed to $\omega_{RF} = 22$ MHz and variable RF amplitude within the range [0…3] mT in 300 steps. The linear combination of clusters with the same $\eta_{\Delta}$ values produces the data set shown in Fig. 5c.